\documentclass[manuscript]{aastex62}

\usepackage{amsmath}
\usepackage{textcomp}
\usepackage{ulem}
\usepackage{graphicx}

\usepackage{hyperref}
\usepackage[hyphenbreaks]{breakurl}

\newcommand{\ps}{\ {\rm s}^{-1}}
\newcommand{\yr}{\ {\rm yr}}

\newcommand{\cm}{\ {\rm cm}}

\newcommand{\tev}{\ {\rm TeV}}
\newcommand{\gevc}{ \ {\rm GeV}\ c^{-1} }
\newcommand{\tevc}{ \ {\rm TeV}\ c^{-1} }
\newcommand{\uG}{\ \mu{\rm G}}
\newcommand{\um}{\ \mu m}

\newcommand{\gray}{$\gamma$-ray}
\newcommand{\grays}{$\gamma$-rays}

\newcommand{\Bsnr}{B_{\rm SNR}}

\newcommand{\Bsnro}{^1B_{\rm SNR} }
\newcommand{\Bsnrt}{{^2B}_{\rm SNR} }
\newcommand{\weo}{^1W_e}
\newcommand{\wet}{^2W_e}
\newcommand{\wpo}{^1W_p}
\newcommand{\wpt}{^2W_p}
\newcommand{\nto}{^1n_t}
\newcommand{\ntt}{^2n_t}

\newcommand{\pceo}{^1p_{c,e} }
\newcommand{\pcet}{^2p_{c,e} }
\newcommand{\so}{^1\alpha}
\newcommand{\st}{^2\alpha}
\newcommand{\plcpt}{^2p_{lc}}

\newcommand{\chandra}{{\sl Chandra}}

\newcommand{\fermi}{{\sl Fermi}}

\received{xxx}
\revised{xx.xx}
\accepted{xx.xx}

\submitjournal{xxx}

\shorttitle{Is Cas A a PeVatron?}
\shortauthors{Zhang \& Liu}

\begin{document}


\title{Is Supernova Remnant Cassiopeia A a PeVatron?}

\email{liusm@pmo.ac.cn; xiaozhang@nju.edu.cn}

\author[0000-0002-9392-547X]{Xiao Zhang}
\affil{School of Astronomy \& Space Science, Nanjing University, 163 Xinlin Avenue, Nanjing~210023, China}
\affil{Key Laboratory of Modern Astronomy and Astrophysics, Nanjing University, Ministry of Education, Nanjing~210023, China}

\author[0000-0003-1039-9521]{Siming Liu}
\affil{Key Laboratory of Dark Matter and Space Astronomy, Purple Mountain Observatory, Chinese Academy of Sciences, 8 Yuanhua Road, Nanjing 210034, China}

\begin{abstract}

Cassiopeia A, a well-observed young core-collapse supernova remnant (SNR), is considered as one of the best candidates for studying very high-energy particle acceleration up to PeV via the diffusive shock mechanism. 
Recently, MAGIC observations revealed a \gray\ spectral cutoff at $\sim3.5\tev$, suggesting that if the TeV \grays\ have a hadronic origin, SNR Cas A can only accelerate particles to tens of TeV.
Here, we propose a two-zone emission model for regions associated with the forward (zone 1) and inward/reverse shocks (zone 2). 
Given the low density in zone 1, it dominates the high-frequency radio emission, soft X-ray rim via the synchrotron process and TeV \gray\ via the inverse Comptonization.
With a relatively softer particle distribution and a higher cut-off energy for electrons, emissions from zone 2 dominate the low-frequency radio, hard X-ray via the synchrotron process and GeV \gray\ via hadronic processes.
There is no evidence for high-energy cutoffs in the proton distributions implying that Cas A can still be a PeVatron.
Hadronic processes from zone 1 dominate very high-energy gamma-ray emission. Future observations in hundreds of TeV range can test this model.

\end{abstract}


\section{Introduction}

Supernova remnants (SNRs) have been considered as the dominant sources of Galactic cosmic rays (CRs, mainly protons) below the ``knee" energy of $\sim3 \times 10^{15}$ eV. 
Charged particles can be accelerated at SNR shocks through the diffusive shock mechanism \citep[e.g.,][and reference therein]{Blandford1987} and may even reach $\sim$ PeV at the early Sedov phase \citep[e.g.,][]{Gabici2009,Ohira2012}.
The detection of a characteristic $\pi^0$-decay \gray\ spectral ``bump" in SNRs interacting with molecular clouds, such as IC 443 \citep{Ackermann2013}, W44 \citep{W44.AGILE.2011,Ackermann2013,W44.AGILE.2014} and W51C \citep{W51C.Fermi.2016}, shows that SNR shocks indeed can accelerate ions to the relativistic regime.
However the lack of evidence for PeV particles in SNRs pose a challenge to this paradigm.

Cassiopeia A (Cas~A, G111.7$-$2.1), the youngest Galactic remnant of a core-collapse supernova, has long been considered as the best candidates for very high-energy CR acceleration in this SNR paradigm for Galactic CRs\footnote{E.g., \url{https://www-zeuthen.desy.de/students/2017/Summerstudents2017/reports/Olena_Tkachenko.pdf}}.
With an estimated age of $\sim 340$ yr \citep{Fesen2006}, it was suggested that the ejecta of Cas A is more or less freely propagating in a circumstellar medium produced by its progenitor winds \citep{Chevalier2003,Vink2004.CasA}
except for very localized encounter with dense clouds associated with X-ray filaments \citep{Zhou2018.CasA,Sato2018}.
At a distance of 3.4 kpc \citep{Reed1995}, the forward shock with a velocity of $\sim 5000\ {\rm km\ps}$ has reached a radius of 2.5 pc.

In radio bands, Cas A has a shell-type morphology with two obvious features: a bright radio ring with a $\sim$1.7 pc radius and a faint outer plateau extending up to $\sim$2.5 pc \citep[][and references therein]{Zirakashvili2014}.
The mean spectral index for the entire remnant is about 0.77 \citep{Baars1977,Rosenberg1970,Anderson1991,Anderson1996}
and the overall radio spectrum hardens toward high frequencies, giving rise to a concaved spectrum dominated by the inner ring and outer plateau at the low and the high frequency, respectively. 
In X-ray band, it has been extensively studied, showing an outer thin rim and a bright diffuse inner ring \citep[e.g.,][]{Gotthelf2001,Vink2003.CasA,Hwang2004,Bamba2005,Araya2010a}. 
NuSTAR observations show that the nonthermal hard X-ray emission is dominated by a few hot spots near the projected center of the remnant, which is distinct from nonthermal soft X-ray emission associated with the forward shocks \citep{Grefenstette2015}.
Recently, it has been shown that these hot spots are associated with very fast inward-moving shocks \citep{Sato2018}.
Using 10-year data from INTEGRAL observations, \citet{Wang2016.L} first extended the nonthermal X-ray spectrum to $\sim$220 keV without evidence for a high-energy cutoff.
In GeV band, it was detected by {\sl Fermi}-LAT as a point source with a photon index of 1.9--2.4 \citep{Cas-A.Fermi.2010,Cas-A.Fermi.2013,Cas-A.Fermi.2014} and a spectral break around 1.7 GeV was reported \citep{Cas-A.Fermi.2013,Cas-A.MAGIC.2017}.
As a TeV \gray\ source, it was first detected by HEGRA at 5$\sigma$ level \citep{Cas-A.HEGRA.2001} and was further observed by MAGIC \citep{Cas-A.MAGIC.2007} and VERITAS \citep{Cas-A.VERITAS.2010}.

Two zone emission models have been proposed for the radio to X-ray morphology and spectra \citep{Atoyan2000.TA,Atoyan2000.AT} and the nature of the \gray\ emission is still ambiguous \citep[e.g.,][]{Araya2010b,Cas-A.Fermi.2013,Cas-A.Fermi.2014,Zirakashvili2014}. 
With 158h of high quality data, a high-energy cutoff of $\sim3.5\tev$ in the \gray\ spectrum of Cas~A was recently detected by MAGIC with 4.6$\sigma$ significance \citep{Cas-A.MAGIC.2017}.
This spectral feature seems to disfavor Cas A as a PeV particle accelerator if the TeV \gray\ emission is dominated by hadronic processes.

In this paper, we propose a two-zone model where forward shocks produce a relatively harder high-energy particle distribution than the inward/reverse shocks.
Given the low density associated with the forward shocks, it dominates high-frequency radio emission, soft X-ray rim via the synchrotron process and TeV \gray\ via the inverse Comptonization.
The reverse shocks are associated with a high density zone. With a relatively softer particle distribution and a higher cut-off energy for electrons, it dominates low-frequency radio, hard X-ray via the synchrotron process and GeV \gray\ via the hadronic processes.
There is no evidence for high-energy cutoffs in the proton distributions implying that Cas A can still be a PeVatron. 
The model description and spectral fit are given in Section \ref{sec:mod}.
In section 3, we discuss implications of the model and draw conclusions.

\section{A two-zone model}\label{sec:mod}

\subsection{Model description}
Based on the two-zone model proposed by \citet{Atoyan2000.TA} and features in X-ray band, we treat the outer ``thin rim" (the forward shocks) and the diffuse region immediately behind the forward shock as zone 1.
Zone 2 then includes the rest of emission regions (mainly the bright radio ring, knots, and interior, especially, regions containing inward-moving shocks).
For the sake of simplicity, we assume that the distribution of accelerated particles in each emission zone has a power-law form with a high-energy cutoff in momentum space,
\begin{equation}\label{eq:dis}
N(p) = A\cdot p^{-\alpha}\ {\rm exp}(-p/p_{c})
\end{equation}
where $p$ is the particle momentum, $p_c$ is the cutoff momentum, $\alpha$ is the power-law index for electrons and protons in a given emission zone, and $A$ is the normalization\footnote{In the following, subscripts `e' and `p' will be used for parameters of electrons and protons, respectively and superscripts `1' and `2' denote parameters of mission zone 1 and 2, respectively.} which will be replaced by the total energy of particles with momentum above 1 $\gevc$, $W$, below.
To include the radiative cooling effect on the accumulated electron distributions in these emission zones \citep{Heavens1987,Zirakashvili2010}, we assume that the spectral indexes of these electron distributions increase by 1 above a break energy $E_{bre}=4\ (B_{\rm SNR}/100\uG)^{-2}(t/340\yr)^{-1}\tev$, where $B_{\rm SNR}$ and $t$ are the magnetic field and the shock age, respectively. We also consider the cases where the electron distributions have a super-exponential cutoff of ${\rm exp}(-p^2/p_c^2)$. For zone 1, the shock age will be the same as the age of the remnant. For zone 2, the age of the inward shocks is treated as a free parameter.

To fit the broad-band spectrum, we consider four radiation processes: synchrotron emission from energetic electrons traversing a magnetic field; inverse Compton (IC) scattering on soft seed photons by relativistic electrons;
bremsstrahlung by relativistic electrons;
and \grays\ produced via decay of $\pi^{0}$ due to inelastic collisions between relativistic protons and nuclei in the backgroud.
The photon emissivity for synchrotron, bremsstrahlung and IC processes given in \citet{Blumenthal1970} are adopted. 
For the IC process, the seed photon fields include the cosmic microwave background (CMB) and a far infrared field with a temperature of 100 K and an energy density of 2 ${\rm eV\ cm^{-3}}$ \citep{Mezger1986}.

For the p-p process, we use the analytic photon emissivity developed by \citet{Kelner2006}, including an enhancement factor of 1.84 due to contribution from heavy nuclei \citep{Mori2009}. The density of target gas in zone 1, $\nto$, is 4 times the ambient gas density upstream of the forward shock \citep[$n_0=0.9 \cm^{-3}$,][]{Lee2014.CasA} for a strong shock.
The mean gas density $\ntt$ in zone 2 can be estimated with the supernova ejecta mass, $\ntt=10 \cm^{-3}$ \citep[e.g.,][]{Araya2010b}.
In addition, the cutoff momentum for protons in both zones is fixed at $p_{c,p}=3\ {\rm PeV}\ c^{-1}$, corresponding to the ``knee" energy.

\subsection{Results}

\begin{figure}[htb]
\centering
\includegraphics[height=60mm,angle=0]{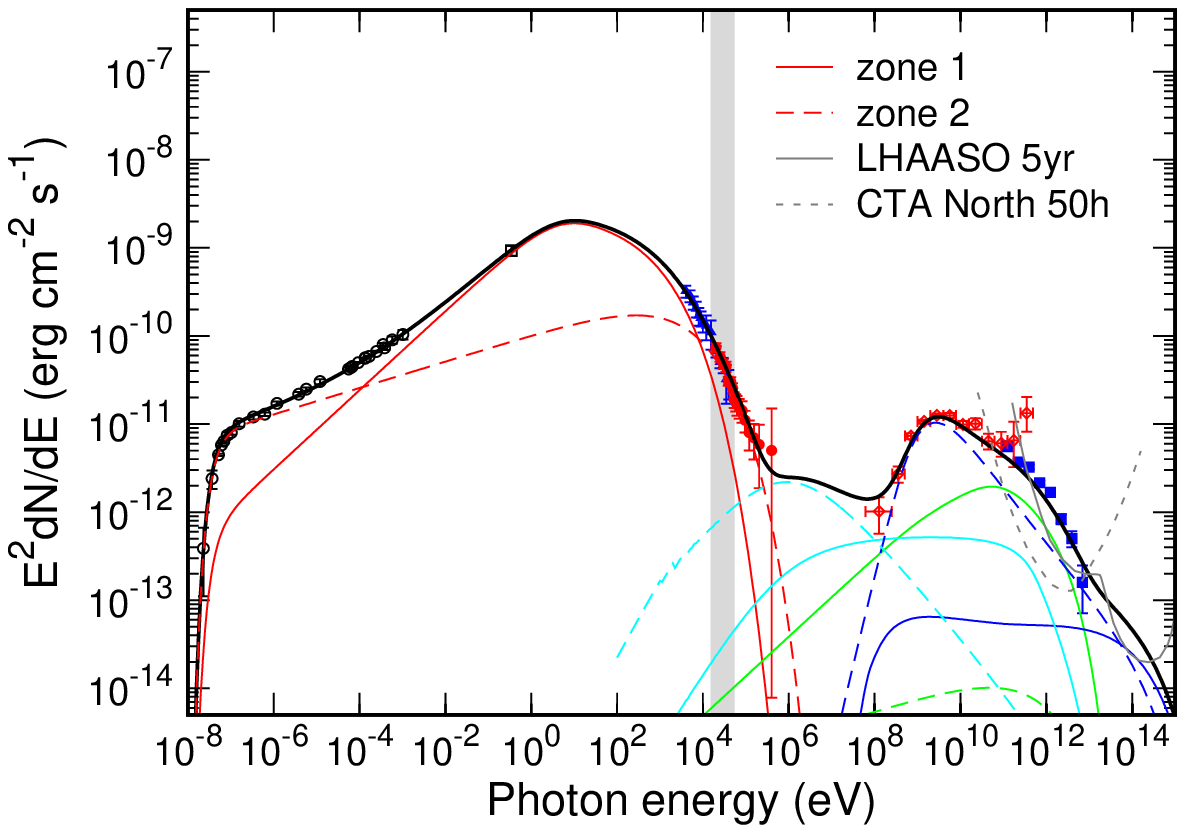}
\includegraphics[height=60mm,angle=0]{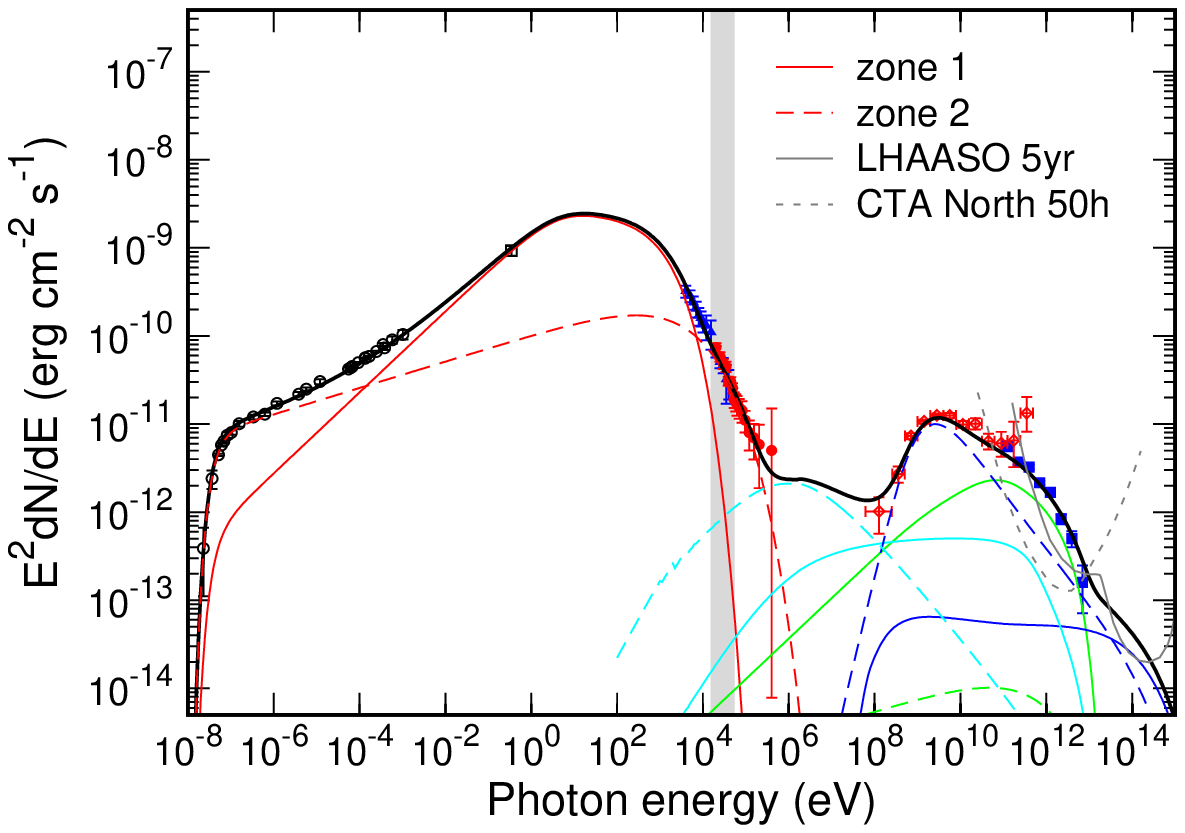}
\caption{
Eye-fitting SED of SNR Cas~A. The black solid line represents the total emission from zone 1 (solid) and 2 (dashed) with various components considered in this work: synchrotron (red), inverse Compton (green), bremsstrahlung (cyan) and p-p collision (blue). Also shown are the radio data (open circle) given in \citet{Vinyaikin2014}, infrared data from IRAC 3.6 $\um$ \citep[open square;][]{DeLooze2017}, X-ray data from Suzaku \citep[filled triangle;][]{Maeda2009} and INTEGRAL-IBIS \citep[filled circle;][]{Wang2016.L}, $\gamma$-ray data from \fermi-LAT and MAGIC \citep[open diamond and filled square, respectively;][]{Cas-A.MAGIC.2017}.
The gray region represents the energy range 15--55 keV. The model parameters (Model A) are given in Table~\ref{tab:param}. 
The sensitivities of LHAASO (gray solid line) and CTA (gray dashed line) are also displayed. The left and right panels show the cases with a normal and super-exponential cutoff in zone 1, respectively. See text for details.
}
\label{fig:sed}
\end{figure}

\begin{figure}[htb]
\centering
\includegraphics[height=60mm,angle=0]{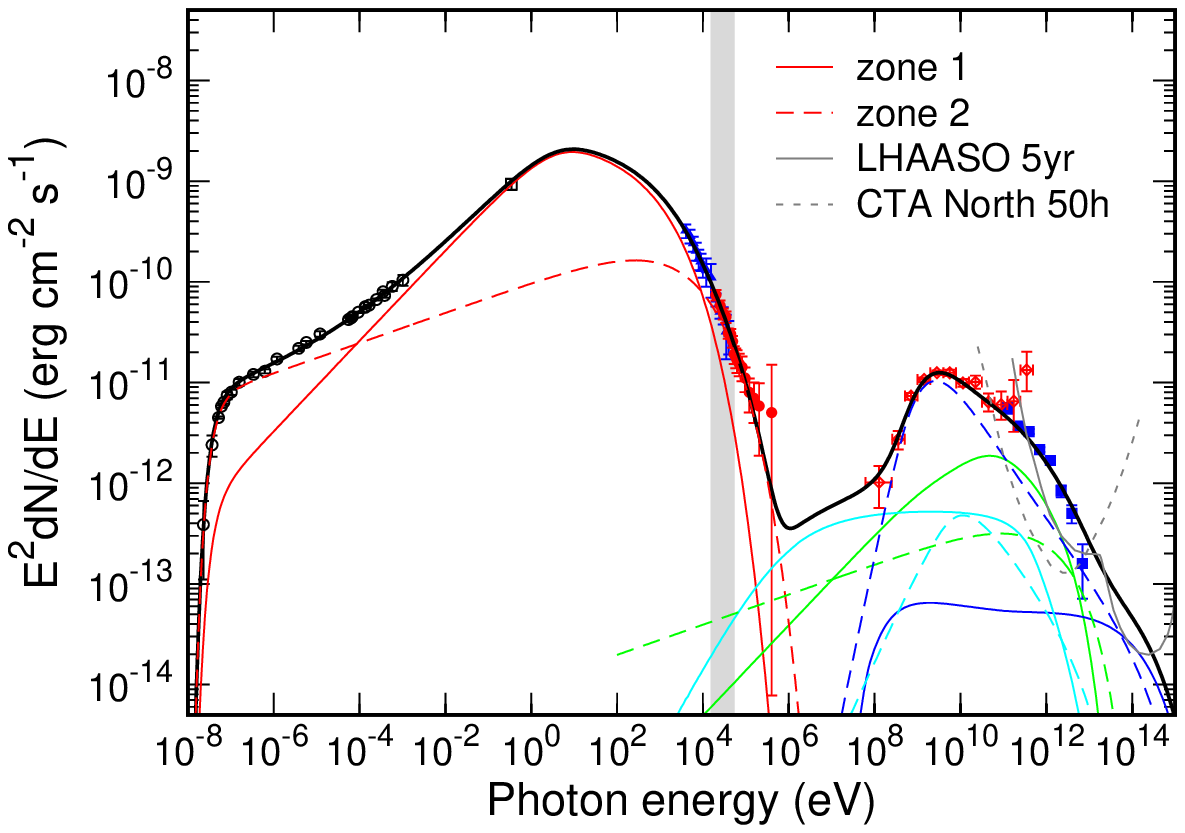}
\includegraphics[height=60mm,angle=0]{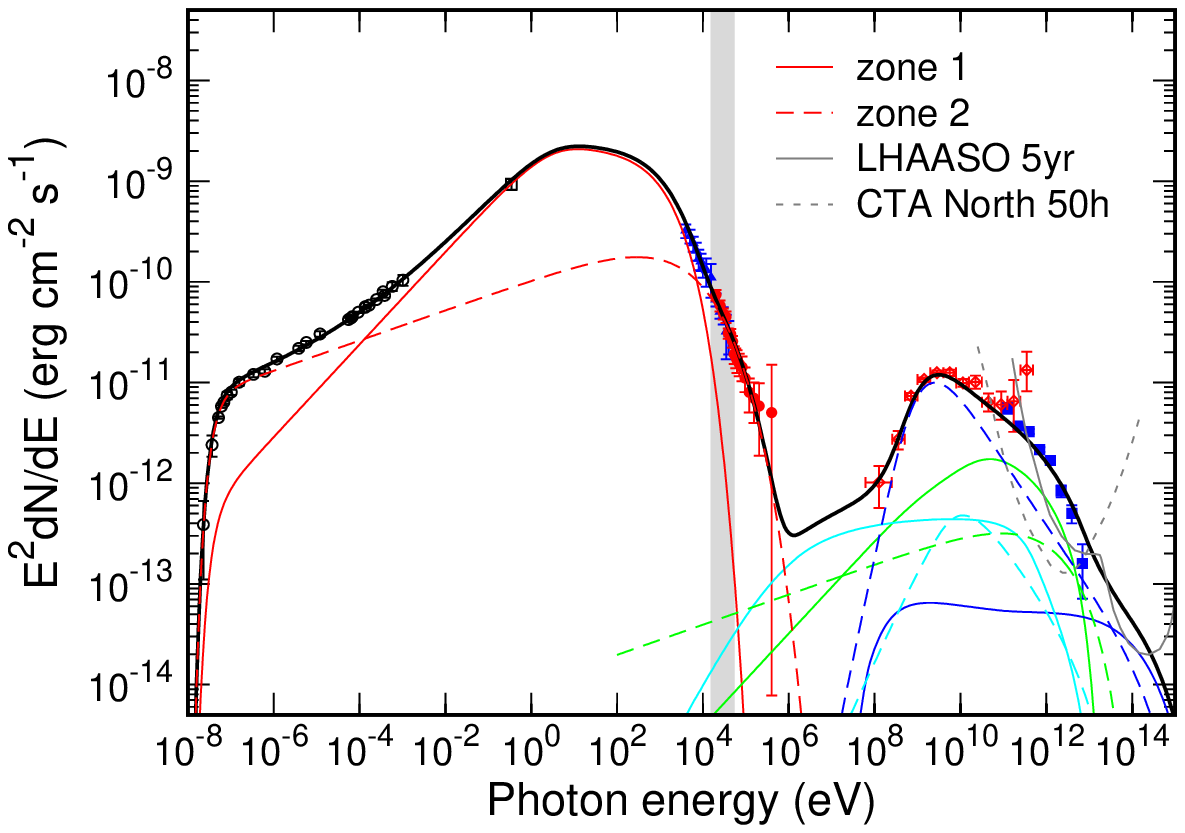}
\caption{
Same as Figure~\ref{fig:sed} but for Model B.
}
\label{fig:sed_b}
\end{figure}

\begin{figure}[htb]
\centering
\includegraphics[height=60mm,angle=0]{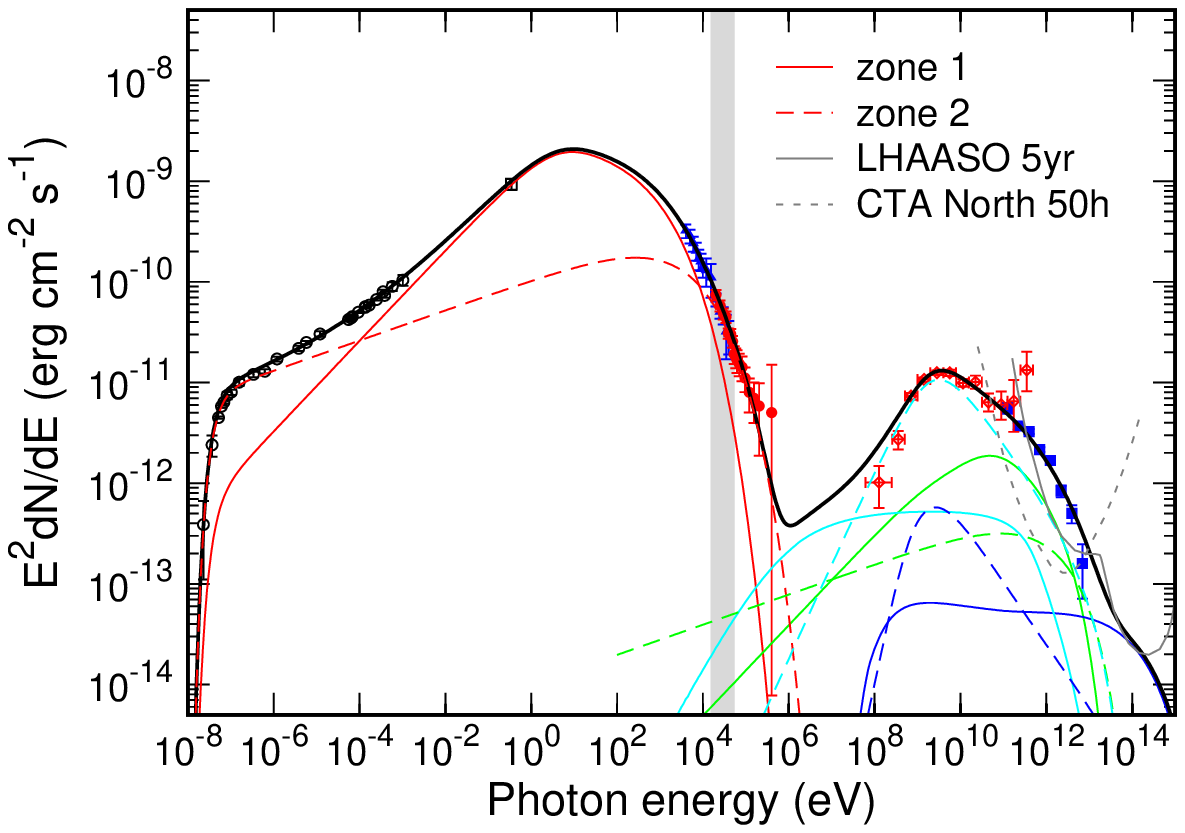}
\includegraphics[height=60mm,angle=0]{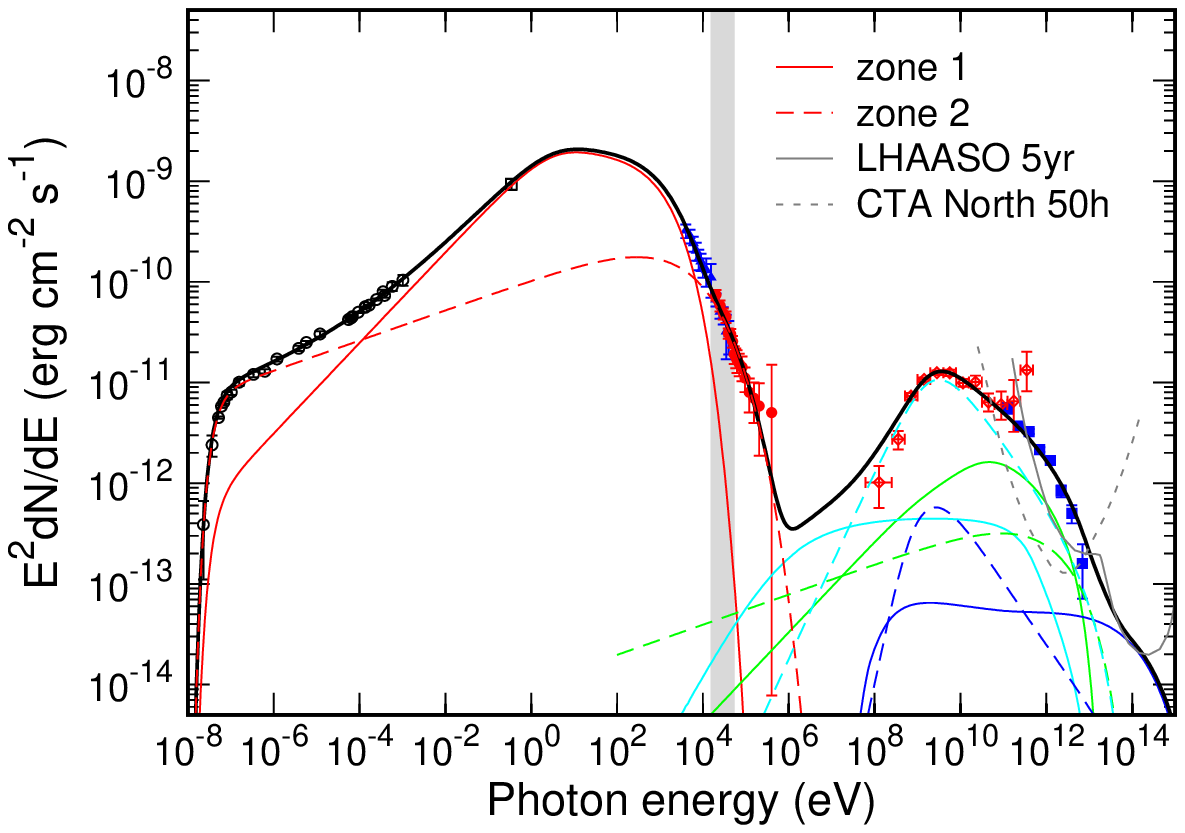}
\caption{
Same as Figure~\ref{fig:sed} but for Model C.
}
\label{fig:sed_c}
\end{figure}

\begin{figure}[htb]
\centering
\includegraphics[height=60mm,angle=0]{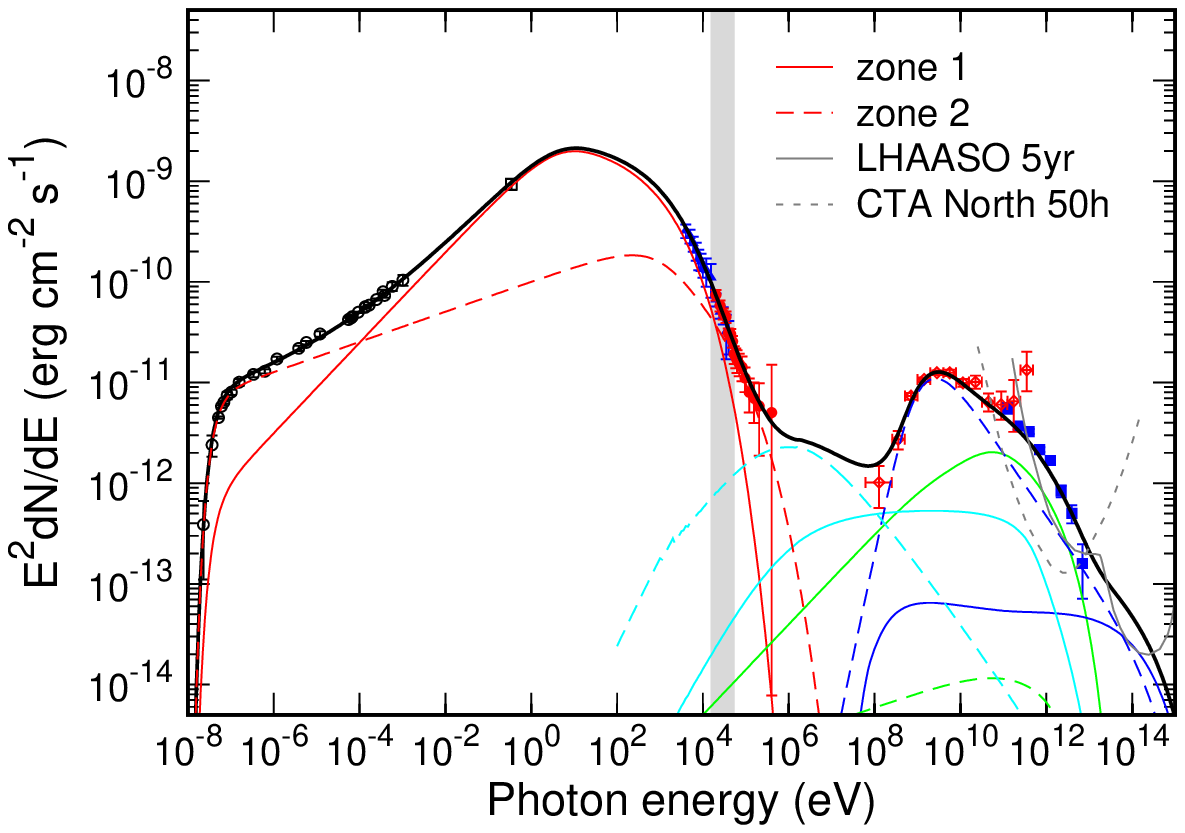}
\includegraphics[height=60mm,angle=0]{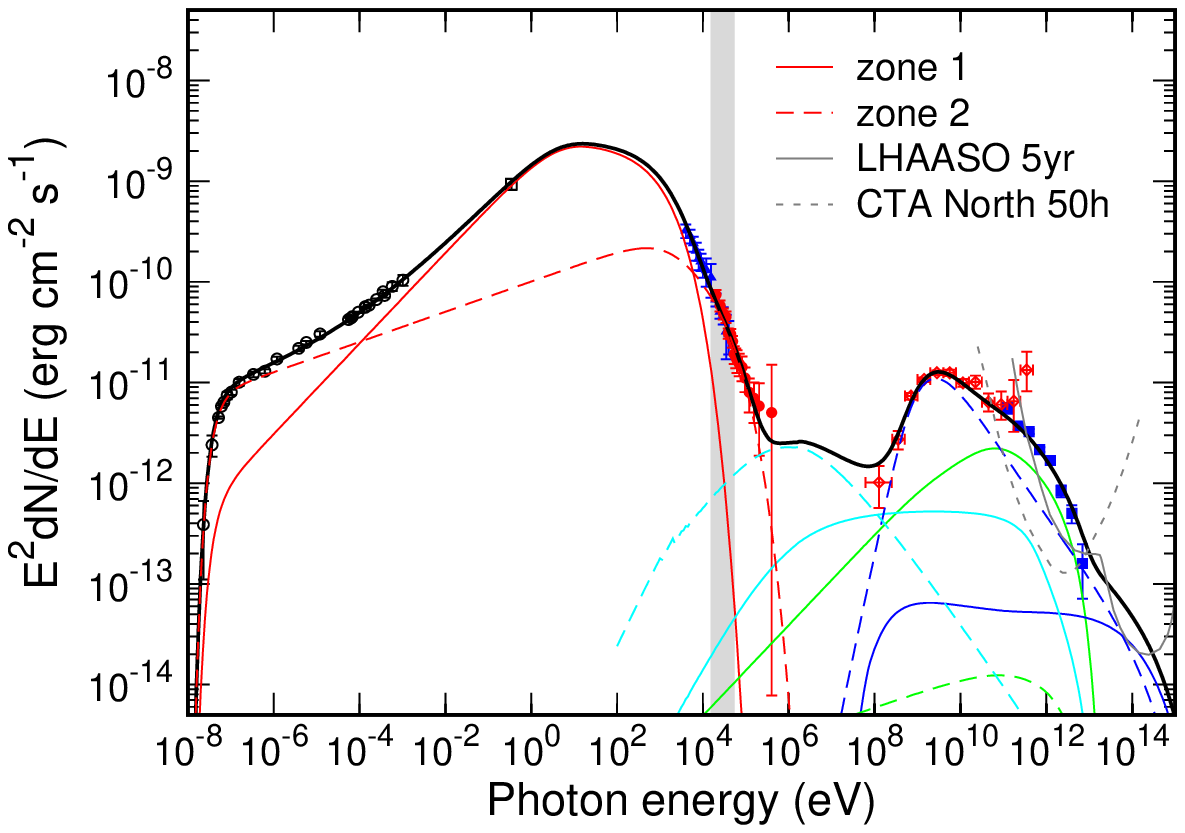}
\caption{
Same as Figure~\ref{fig:sed} but for Model D. The age of the inward shocks is 5 (left) and 3 (right) years.
}
\label{fig:sed_d}
\end{figure}

To start with, we assume that inward shocks in zone 2 form continuously so that the electron distribution may be described by equation~\ref{eq:dis}.
There are therefore five parameters for each emission zone.
The eye-fitting spectral energy distribution (SED) is shown in Figure~\ref{fig:sed} and the corresponding parameters are summarized in Table~\ref{tab:param} (Model A).

In the \gray\ band, the GeV emission is dominated by the p-p process in zone 2 (blue dashed line), while the TeV emission mainly comes from the IC process in zone 1 (green solid line).
The corresponding energy content in these relativistic particles are $\weo=7.0\times10^{47}$~erg and $\wpt=1.8\times10^{50}(\ntt/10\cm^{-3})^{-1}$ erg, and $\so = 2.1$ and $\st=2.7$.
The magnetic field in zone 1 should be considered as a lower limit. Otherwise one would expect TeV emission exceeding the observed flux level.
The electron cutoff momentum in zone 1 is well-constrained by the soft X-ray and TeV \gray\ spectra: $\pceo=7.0\tevc$, that is slightly lower than $\pcet=9.0\tevc$ in zone 2 constrained mostly by the hard X-ray spectrum.

\citet{Sato2018} recently showed that hard X-ray emission of Cas~A is associated with fast inward-moving shocks with a speed of $\sim5100-8700\ {\rm km\ s^{-1}}$, which is higher than the speed of $\sim5000\ {\rm km\ s^{-1}}$ for forward shocks.
Considering feedback of accelerated particles to the shock structure, \citet{Bell2004} argued that the cutoff momentum increases with the shock speed, which is consistent with values of the cutoff momentums in these two zones derived from our spectral fit.
Cas~A therefore is unique in the sense that there are inward shocks that accelerate particles to higher energies than forward shocks \citep{Telezhinsky2012}.

The GeV spectrum of Cas~A also has a distinct cutoff below $\sim$2 GeV.
To explain this spectral feature, a low-energy cutoff $\plcpt=15\gevc$ is introduced to the proton distribution\footnote{When calculating the total energy $W_p$, we do not consider this low-energy cutoff effect.}.
The low flux level at $\sim$100 MeV also leads to an upper limit for $\wet<4\times10^{46}$ erg for the bremsstrahlung process of relativistic electrons\footnote{We assume that the electron distribution cuts off at 1 MeV at low energies.}, which implies a lower limit to $\Bsnrt>1.0$~mG for a given synchrotron spectrum.

If we attribute the low-energy cutoff of the proton distribution to slower diffusion of lower energy particles into high-density emission regions, the bremsstrahlung emission of relativistic electrons is also subject to the same process.
We therefore expect that for the bremsstrahlung process, the low energy cutoff of relativistic electrons should be the same as that for protons (Model B).
The right panel of Figure~\ref{fig:sed_b} shows the fitted results for such a scenario with $\Bsnrt=160\uG$ and $\wet =1.0\times10^{48}$~erg.
The hundreds of keV to 100 MeV spectrum is dominated by electron bremsstrahlung in zone 1 and is distinct from that for Model A where electron bremsstrahlung in zone 2 dominates.
An even lower value for $\Bsnrt$ will lead to a 100 MeV \gray\ flux exceeding the observed value for the IC of zone 2.

If the target of the electron bremsstrahlung and proton processes in zone 2 has a much higher density as indicated by the low cutoff energy discussed above, $\Bsnrt$ will have a much higher value if the GeV emission is still dominated by hadronic processes. Moreover, the GeV emission may also be dominated by the electron bremsstrahlung (Model C).
The right panel of Figure~\ref{fig:sed_c}, shows such a spectral fit with $\ntt=100\cm\ps$ and $\plcpt=4\gevc$.
It slightly overproduces \gray\ below $\sim$2 GeV and predicts very low fluxes above 10 TeV.
For an even higher density of the emission region, both $\wet$ and $\wpt$ will be less than $1.0\times 10^{48}$ erg and $\Bsnrt$ greater than 160 $\uG$. However, $^2W_{e/p}$ should be considered as a lower limit for accelerated particles since some of them haven't reached the high-density emission regions.

Next, we assume that inward shocks formed instantaneously with the age $t$ as another free parameter.
Figure~\ref{fig:sed_d} shows a spectral fit similar to Model A (Model D) with an inward shock age of 5 and 3 years for the left and right panels, respectively, implying very efficient acceleration at inward shocks, which is reasonable for the high shock speeds and strong magnetic field.
The shock age can be longer for models similar to Models B and C for a weaker magnetic field.
However, we found that for the soft electron distribution in zone 2, models with a low break energy can not reproduce the observed hard X-ray fluxes via synchrotron process.

\begin{figure}
\centering
\includegraphics[height=60mm,angle=0]{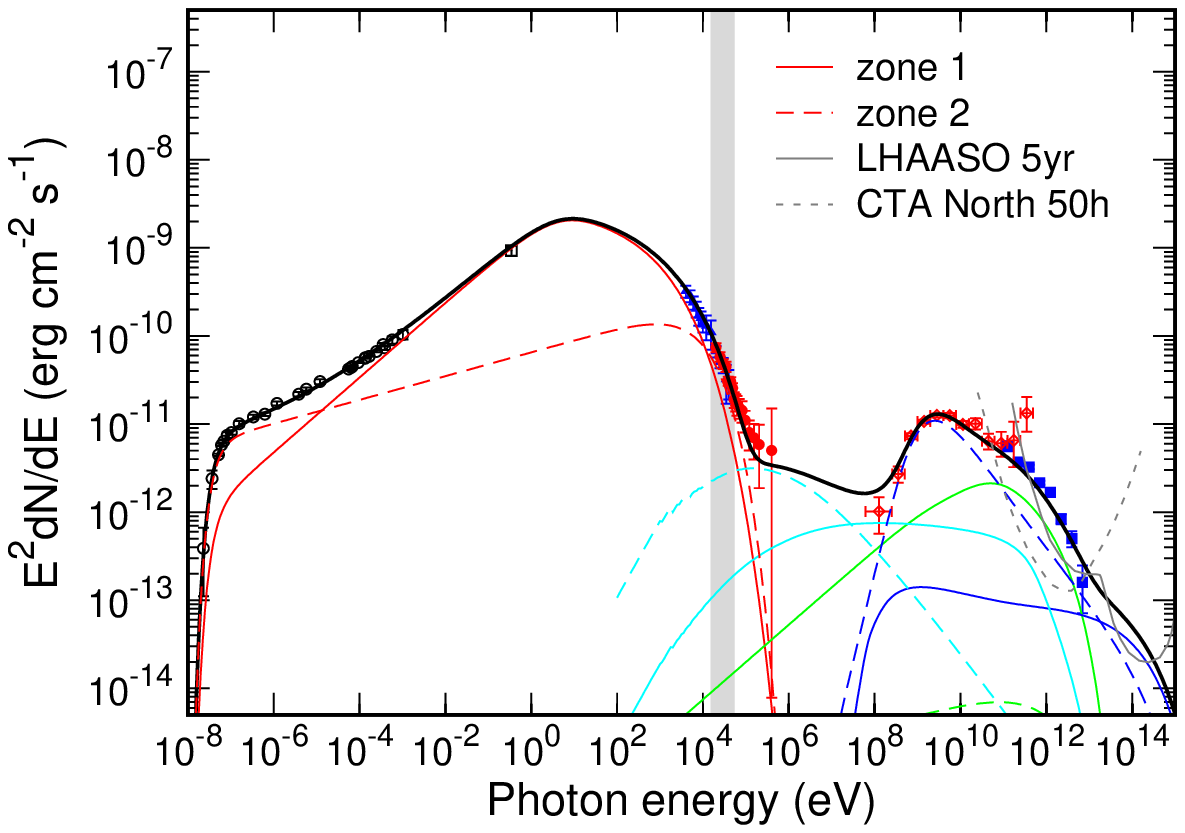}
\includegraphics[height=60mm,angle=0]{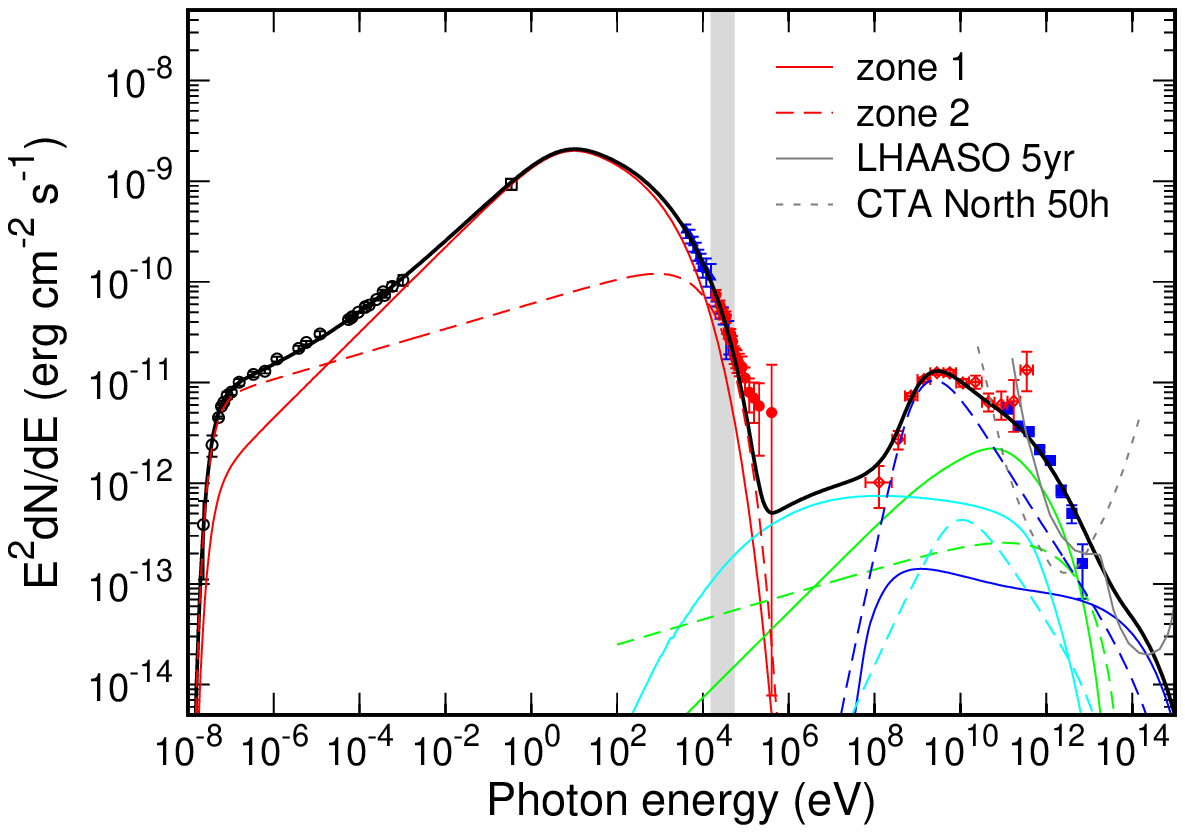}
\includegraphics[height=60mm,angle=0]{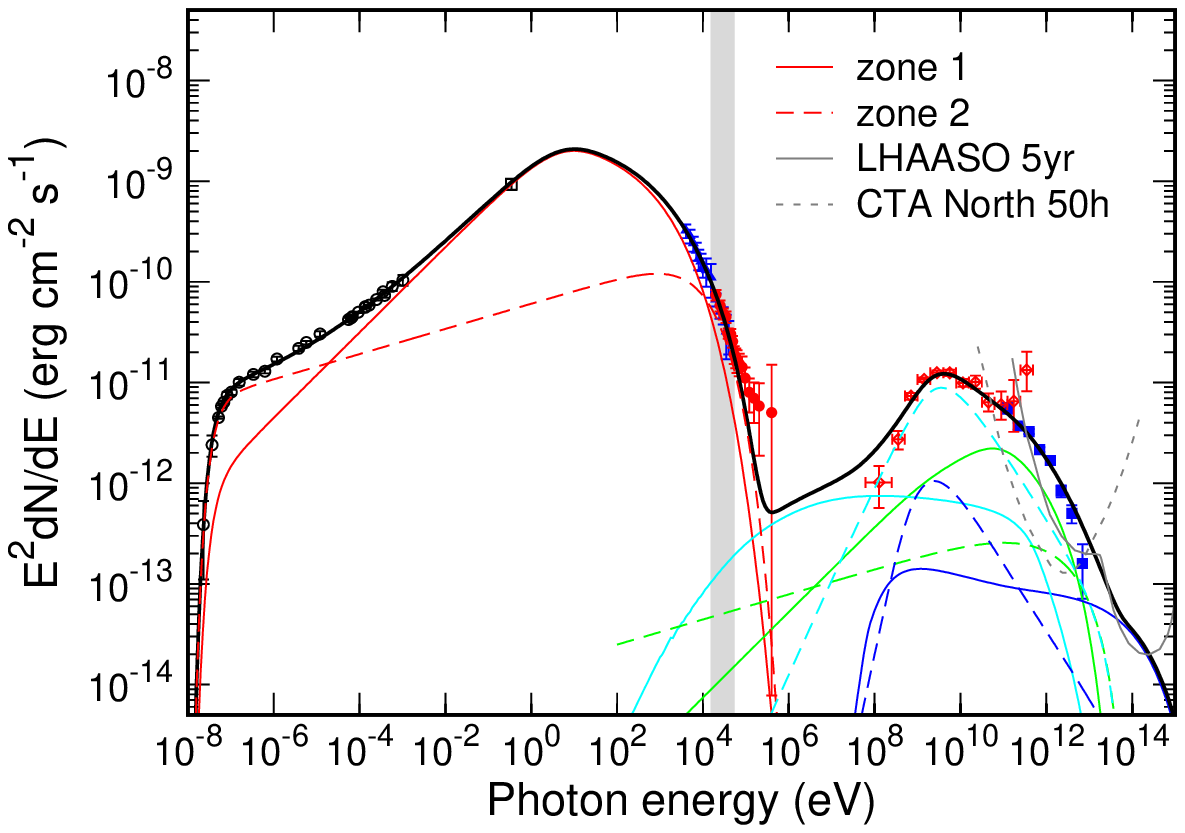}
\includegraphics[height=60mm,angle=0]{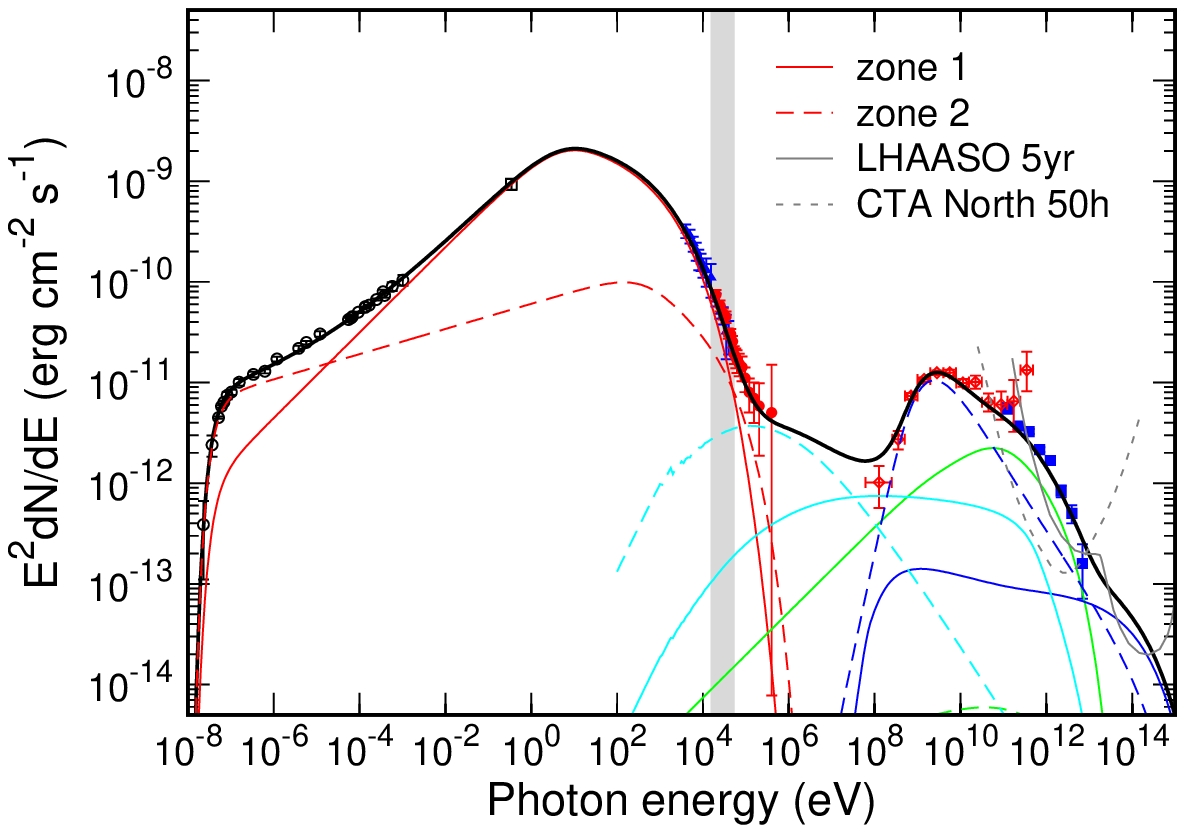}
\caption{
Same as Figure~\ref{fig:sed} but with a super-exponential cutoff in zone 2. Model parameters are given in Table 2. A: Top-left; B: Top-right; C: Bottom-left; D: Bottom-right.
}
\label{fig:sed_abcd}
\end{figure}

To maximize contributions to hard X-ray from zone 1, we also consider models with a super-exponential cutoff in zone 2 and exponential cutoff in zone 1.
Figure~\ref{fig:sed_abcd} shows the spectral fits with the corresponding model parameters given in Table~\ref{tab:param2}.
For Model D, the age of the inward shocks is 5 years. Here to fit the hard X-ray spectrum, the low cutoff energy of the electron distribution has been reduced to 500 keV.
However, Models B and C under-produces the overall hard X-ray fluxes significantly.
A better treatment of bremsstrahlung emission from zone 1 at non-relativistic energies may address this issue and is beyond the scope of this paper.

The \gray\ emission via hadronic processes from zone 1 dominate the hundreds of TeV range for $\wpo=1.0\times10^{48}$ erg, which is also sufficient to account for the high-energy component of CR ion spectra in the SNR paradigm for Galactic CRs \citep{Zhang2017.LY}.
Future observations in this energy range will be able to distinguish these models.
Models with a super-exponential high-energy cutoff in zone 1 have a softer X-ray spectrum and weaker hard X-ray emission in zone 1 than these with an exponential high-energy cutoff.
Detailed analysis of non thermal X-ray emission from different emission zones is warranted.


In general, the model parameters for zone 1 are well-constrained, and we note that the acceleration efficiencies of electrons and protons are comparable for the forward shocks, which appears to be in conflict with the observed CR electrons and protons fluxes.
Recent modeling of CR ion spectra by \citet{Zhang2017.LY} shows that young SNRs may dominate acceleration of high-energy CRs while the bulk CRs at lower energies are accelerated by slow shocks in old SNRs.
We therefore expect a more prominent high-energy component in the CR electron spectrum.
The modeling of CR electron spectrum by \citet{Li2015.electron} indeed reveals such a high-energy component.
Considering the efficient radiative loss of high-energy electrons in the Galaxy, their flux relative to protons at the Earth should be much lower than that in their source of origin.
The relative acceleration efficiencies of electrons and protons in zone 2 are not well constrained. Although there are efficient ion acceleration for Models A and B, the spectrum is soft and these particles are also subject to adiabatic loss as the remnant expands, the observed high-energy CR is likely dominated by particles accelerated in zone 1.

The synchrotron emission from electrons in zone 1 dominates the soft X-rays and high-frequency radio fluxes (red solid line), 
which, in combination with the TeV fluxes dominated by the IC process, constrains the magnetic field in zone 1 $\Bsnro=250\uG$.
The hard X-rays and low-frequency radio data can be explained by synchrotron emission from zone 2 (red dashed line) for the relatively softer particle distribution and higher cutoff energy.
The magnetic field in zone 2 is larger than $160\uG$ and the
low frequency radio spectrum is also cut off via an absorption term $e^{-\tau_0(\nu/10\ {\rm MHz})^{-2.1}}$ with $\tau_0=0.9$ at 10 MHz.

\begin{deluxetable}{cclllll}[htb]
\tablecaption{Fitted parameters. Parameters in parenthesis correspond to electron distribution with a super-exponential cutoff in emission zone 1. \label{tab:param}}
\tablecolumns{7}
\tablewidth{0pt}
\tablehead{
  \colhead{Model} &
  \colhead{zone} &
  \colhead{$\alpha$} &
  \colhead{$p_{c,e}$} & 
  \colhead{$\Bsnr$} & 
  \colhead{$W_e$} & 
  \colhead{$W_p$} \\
  & 
  & 
  & 
  $({\rm TeV\ c^{-1}})$ & 
  $(\mu$G) & 
  $(10^{47}$ erg) &  
  $(10^{48}$ erg)
}
\startdata
 A & 1 & 2.1 &  7.0 (10.0) &  250 &  7.0 & 1.0 \\
   & 2 & 2.7 &  9.0 & 1000 &  0.4 & 180 \\
\hline
 B & 1 & 2.1 & 7.0 (10.0) &  260 (280) &  7.0 (6.0) & 1.0 \\
   & 2 & 2.7 & 22.0  &  160 & 10.0 & 180 \\
\hline
 C & 1 & 2.1 & 7.0 (10.0) &  260 (280) &  7.0 (6.0) & 1.0 \\
   & 2 & 2.7 & 22.0  &  160 & 10.0 & 1.0 \\
\hline
 D & 1 & 2.1 & 8.0 (10.0) &  250 &  7.0 & 1.0 \\
   & 2 & 2.7 & 25.0  &  950 & 0.4  & 200 \\
\enddata
\end{deluxetable}

\begin{deluxetable}{cclllll}[htb]
\tablecaption{Fitted parameters but with a super-exponential cutoff in zone 2.\label{tab:param2}}
\tablecolumns{7}
\tablewidth{0pt}
\tablehead{
  \colhead{Model} &
  \colhead{zone} &
  \colhead{$\alpha$} &
  \colhead{$p_{c,e}$} & 
  \colhead{$\Bsnr$} & 
  \colhead{$W_e$} & 
  \colhead{$W_p$} \\
  & 
  & 
  & 
  $({\rm TeV\ c^{-1}})$ & 
  $(\mu$G) & 
  $(10^{47}$ erg) &  
  $(10^{48}$ erg)
}
\startdata
 A & 1 & 2.15 &  8.0 &  250 &  9.0  & 2.0 \\
   & 2 & 2.73 & 14.0 & 1000 &  0.25 & 200 \\
\hline
 B & 1 & 2.15 &  8.0 &  240 &  9.0 & 2.0 \\
   & 2 & 2.75 & 40.0 &  140 & 10.0 & 200 \\
\hline
 C & 1 & 2.15 &  8.0 &  240 &  9.0 & 2.0 \\
   & 2 & 2.75 & 40.0 &  140 & 10.0 & 2.0 \\
\hline
 D & 1 & 2.15 &  9.0 &  240 &  9.0 & 2.0 \\
   & 2 & 2.75 & 30.0 & 1000 & 0.25  & 200 \\
\enddata
\end{deluxetable}

\section{Discussion and Conclusion}\label{sec:dc}
Non-thermal emission of Cas A has been extensively studied via radio and X-ray observations.
Recent $\gamma$-ray observations reveal spectral cutoffs at both low- and high- energy ends of the spectrum.
We have shown that a simple two zone emission model can not only account for the radio and X-ray morphology and spectra of Cas A, but also reproduce the \gray\ spectrum with the proton distributions cutting off above 1 PeV. 
In this model, the TeV \grays\ are dominated by the IC process of electrons in zone 1 for forward shocks, giving rise to the observed TeV spectral cutoff. 
Thus, the high-energy cutoff at $\sim 3.5$ TeV detected by MAGIC \citep{Cas-A.MAGIC.2017} does not imply that the proton spectrum in this SNR cuts off at a few tens of TeV and Cas A can still be an efficient PeVatron.

The GeV \grays\ are dominated by the p-p process in zone 2, and the low-energy cutoff in proton spectrum at momentum $\plcpt\approx15\gevc$ may be attributed to slower diffusion of lower energy protons accelerated by the inward/reverse shocks into the high density regions \citep[e.g.,][]{Gabici2014}.
Given the low flux observed at $\sim$100 MeV and the soft particle distribution in zone 2, the electron acceleration is much less efficient than protons and the magnetic field should be stronger than 1 mG (Model A).
If the bremsstrahlung of relativistic electrons in zone 2 is also subject to the slow diffusion of low energy particles into high-density regions (Model B), we predict that the hard X-ray spectrum cuts off near $\sim$100 keV and a new hard spectral component emerges above $\sim$MeV due to electron bremsstrahlung in zone 1.
In such a model the magnetic field has a low limit of $140\uG$.
If the density of the target regions is much higher than the fiducial value of $10 \cm^{-3}$ adopted above (Model C), the GeV \grays\ can be dominated by electron bremsstrahlung in zone 2 as well.
The model produces a relatively softer spectrum below $\sim$2 GeV and very low flux above 10 TeV.
A $\wpo=1.0\times10^{48}$ erg is compatible with current observations and is also high enough to produce the high-energy component of CR ion spectra via young SNRs \citep{Zhang2017.LY}.
Then proton processes in zone 1 dominate gamma-ray emission above ~100 TeV. 
Future $\gamma$-ray observations will be able to distinguish these models.

Our model therefore predicts that most of the GeV emission comes from inner region of the SNR
, which is consistent with \fermi\ results reported by \citet{Cas-A.Fermi.2013} and \citet{Cas-A.Fermi.2014}.
Although more protons are accelerated in zone 2 than in zone 1, contributions to high-energy CRs are likely dominated by forward shocks of zone 1 since protons in zone 2 have a relatively soft distribution and are also subject to adiabatic loss as the remnant expands before escaping into the interstellar medium.

NuSTAR observations show that hard ($>15$ keV) X-ray emission from SNR Cas A is dominated by bright knots near the remnant center \citep{Grefenstette2015}.
Recent X-ray observations reveal that these bright knots are associated with high speed inward-moving shocks \citep{Sato2018}, which is consistent with a higher high-energy cutoff for the electron distribution introduced for this region \citep{Bell2004}. 
At the same time, the synchrotron emission from electrons in zone 1 dominate the non-thermal soft X-rays which is associated with the outer thin rim resolved by \chandra\ \citep[e.g.,][]{Hwang2004,Bamba2005,Araya2010a}.

For the cooling-limit cutoff, we use the normal exponential form (${\rm exp}[-(p/p_c)^{\delta}]$ with $\delta=1$) which correspond to diffusive shock acceleration with an energy-independent diffusion coefficient \citep[e.g.,][]{Zirakashvili2010}.
If the diffusion coefficient is Bohm-like, then the cutoff should be super exponential with $\delta=2$ \citep{Zirakashvili2010}.
For each model, we refit the broadband spectra with a super exponential cutoff in zone 1 and plot the SEDs in the right panel for each figure.
The fitted parameters with values different from that for $\delta=1$ are shown in parenthesis in Table~\ref{tab:param}.

In general models with a super exponential cutoff give a slightly better fit to the TeV data.
Models with a normal exponential cutoff can produce significant hard X-ray emission in zone 1, in agreement with observations \citep{Grefenstette2015}.
Contribution to hard X-ray flux from zone 1 can be maximized by adopting a super-exponential cutoff for zone 2 (Figure~\ref{fig:sed_abcd} and Table~\ref{tab:param2}).
More quantitative analysis of hard X-ray emission from zone 1 and 2 is needed to distinguish these models.

In conclusion, the two-zone model presented in this study explains the spatial and spectral features of Cas A, and predicts that 1) the TeV \grays\ have a leptonic origin and mainly come from the outer shell; 2) Cas A can be a PeV accelerator and may be bright at $>$ 100 TeV band.
These predictions will be tested by the CTA (better angular resolution) and LHAASO (capacity of detecting 100 TeV {\grays}) experiments in the future.

\section*{Acknowledgements}
This work is partially supported by National Key R\&D Program of
China: 2018YFA0404203, NSFC grants: U1738122, 11761131007
and by the International Partnership Program of Chinese Academy
of Sciences, Grant No. 114332KYSB20170008.
X.Z. thanks the support of NSFC grants 11803011, 11851305 and 11633007.

\bibliographystyle{aasjournal}
\bibliography{/Users/xiao/literatures/bib/refs,/Users/xiao/literatures/bib/refs-book,/Users/xiao/literatures/bib/refs-G-ray,/Users/xiao/literatures/bib/refs-obs-HE}


\end{document}